# Data navigation on the ENCODE portal

**Meenakshi S. Kagda, Bonita Lam, Casey Litton, Corinn Small, Cricket A. Sloan, Emma Spragins, Forrest Tanaka, Ian Whaling, Idan Gabdank, Ingrid Youngworth, J. Seth Strattan, Jason Hilton, Jennifer Jou, Jessica Au, Jin-Wook Lee, Kalina Andreeva, Keenan Graham, Khine Lin, Matt Simison, Otto Jolanki, Paul Sud, Pedro Assis, Philip Adenekan, Stuart Miyasato, Weiwei Zhong, Yunhai Luo, Zachary Myers**, J. Michael Cherry and Benjamin C. Hitz

**Corresponding author: Benjamin C. Hitz**
Department of Genetics, Stanford University School of Medicine, Stanford, CA 94305, USA

## Abstract:

Spanning two decades, the Encyclopaedia of DNA Elements (ENCODE) is a collaborative research project that aims to identify all the functional elements in the human and mouse genomes. To best serve the scientific community, all data generated by the consortium is shared through a web-portal (https://www.encodeproject.org/) with no access restrictions. The fourth and final phase of the project added a diverse set of new samples (including those associated with human disease), and a wide range of new assays aimed at detection, characterization and validation of functional genomic elements. The ENCODE data portal hosts results from over 23,000 functional genomics experiments, over 800 functional elements characterization experiments (including *in vivo* transgenic enhancer assays, reporter assays and CRISPR screens) along with over 60,000 results of computational and integrative analyses (including imputations, predictions and genome annotations). The ENCODE Data Coordination Center (DCC) is responsible for development and maintenance of the data portal, along with the implementation and utilisation of the ENCODE uniform processing pipelines to generate uniformly processed data. Here we report recent updates to the data portal. Specifically, we have completely redesigned the home page, improved search interface, added several new pages to highlight collections of biologically related data (deeply profiled cell lines, immune cells, Alzheimer's Disease, RNA-Protein interactions, degron matrix and a matrix of experiments organised by human donors), added single-cell experiments, and enhanced the cart interface for visualisation and download of user-selected datasets. Along with this, we also provide a brief summary of data generated using the ENCODE uniform processing pipelines. Finally we have added extensive help pages, tutorials and documentation to assist researchers from a variety of backgrounds in exploring the universe of ENCODE data.

## Introduction:

The ENCODE web-portal (https://www.encodeproject.org) hosts a rich and diverse collection of biological datasets that are a result of a tremendous NHGRI-funded research consortium level effort that lasted for more than two decades. The project was divided into four phases and the final phase (ENCODE 4) came to a completion in December 2022. During these different phases, research labs across the globe joined forces to handle an enormous task of identifying all functional DNA elements present in the human genome by utilising several novel research approaches and cutting edge techniques in the fields of Genetic Engineering, Genomics, Bioinformatics and Machine Learning. The raw sequencing data and processed results generated by the participating labs were deposited at a central repository and a database that is currently hosted at the ENCODE Data Portal. The portal not only facilitates cross-collaboration within the consortium but also serves as a rich open-source community resource for millions of scientists across the globe.

The current ENCODE data portal was first released in August 2014[1] and has since been expanded to accommodate a variety of consortium requirements. The database currently hosts more than 1.3 PB worth of genomics data with an average traffic of ~25,000 unique users per month world-wide (a total of ~1,300,000 users per year). The portal provides rich metadata for each dataset, along with concomitant documentation. In addition, the user experience is enhanced by an intuitive interface that allows researchers to a) easily navigate the data portal, b) sub-select datasets of their interest using faceted search[2], c) visualise the outputs of multiple genomics datasets by using an in-built Valis genome browser and d) download relevant data using different file selection options. The data portal can also be queried using an API (Application Programming Interface); this ensures that the data can be downloaded using computational methods in addition to web browser based options.

Subsequent to our prior publication in 2020[3], new features have been added to the website. These features have been aimed to improve aspects of user experience as well as to accommodate newer data types. For example, we have completely re-designed the home page such that it provides quick links to several important datasets and collection pages using a single mouse click on the various homepage shortcut cards. We have improved the cart functionality such that it allows creation of curated groupings in several different carts, name them as needed, provide useful descriptions, visualise the bed and bigWig files on a genome browser, filter the results based on both file properties as well as dataset properties within the cart and ultimately download specific file subsets from within the carts. There is also a new Valis genome browser interface called the Encyclopaedia Browser that allows simultaneous visualisation of results from genome annotations that were generated using integrative analysis methods constituting the Encyclopaedia of DNA elements.

New genomic data types produced in this last phase of ENCODE now have their own dedicated pages. Some such data types include the Functional Elements characterization assays (*in vivo* transgenic enhancer assays, massively parallel reporter assays and CRISPR screens). In addition, thousands of functional genomics experiments (including ChIP-seq, DNase-seq, ATAC-seq, RNA-seq, WGBS, micro RNA-seq, long read RNA-seq and HiC) submitted to the ENCODE data portal have been uniformly processed using the latest ENCODE uniform processing pipelines, and the results along with useful summary statistics and quality metrics are made available on the ENCODE portal[4,5].

## Summary of ENCODE datasets and objects:

The ENCODE data portal relies on the various metadata objects submitted by the consortium members to display the wide variety of data available on the portal. As of March 2023, the resource contains a total of ~2.5M (precisely 2,534,857) released objects. Of these, about 775K (precisely 775,887) objects represent released file objects that belong to one of the 95,168 released datasets that are publicly available* (see note below). These datasets can be further categorised into several different types. The seven main sub-categories are: 1) functional genomics experiments, 2) functional characterization experiments (including transgenic enhancer experiments), 3) single cell experiments, 4) series, 5) computational and integrative products (annotation datasets) and 6) publications and corresponding data objects.

In our encodeD database system [https://github.com/ENCODE-DCC/encoded], each of these data types (and other relevant metadata) are captured in schema-defined objects. JSON schema is used to keep the properties and links of the objects consistent. In total we have 118 types of objects. A full list, along with their properties, can be found on the ENCODE profiles page at this url: https://www.encodeproject.org/profiles/. For more information on the ENCODE resource and data objects, the readers can refer to ENCODE help pages and our prior publications[2,6].

A "collection" page of several objects of the same type can be viewed on the portal in at least two formats: search view and report view. For example all the experiment objects on the data portal can be viewed on these two rendered UI pages: https://www.encodeproject.org/search/?type=Experiment and https://www.encodeproject.org/report/?type=Experiment. For some key dataset objects, a matrix view is also available. The different views available for each of the objects have several predefined facets located on the left section of the page which allows further downsizing of the search results based on relevant metadata. The facets also serve as important pointers about the variety of data available for selection and they have been grouped into related categories for the main search pages such that each of the main categories could be expanded or collapsed as needed to further view the complete list of facets. For a more detailed description on using facets and query building, readers can refer to the ENCODE help pages (https://www.encodeproject.org/help/getting-started/) or one of our prior publications[2].

*[Note: In general, publicly available objects on the portal have different statuses such as released, archived or revoked (more details in glossary). For the entire manuscript, unless otherwise noted, we will refer to "publicly available" datasets or files as being equivalent to "released" datasets or files only. In addition, the number of each data type described below refers to the number of released objects excluding community submissions (unless otherwise noted) found on the portal as of March 2023].*

## Functional genomics experiments:

The functional genomics experiments on the ENCODE portal are referred to as "Experiment" objects. There are a total of 23,297 released experiments (https://www.encodeproject.org/search/?type=Experiment&status=released&award.project!=community) hosted on the ENCODE portal. Out of these, 17,162 datasets are a result of various phases of ENCODE awards (ENCODE 2, 3, 4, ENCODE2-mouse) and the rest 6135 experiments were either imported from other allied consortium efforts or directly submitted under those awards such as modENCODE and modERN[7]; Genomics of Gene regulation or GGR[8] and Roadmap Epigenomics Mapping Consortium or REMC[9]. The portal also hosts 204 "community" datasets generated by laboratories not associated with one of the above-mentioned consortia. The breadth of genomics datasets (Fig 1) generated by the ENCODE funded labs is evident by the variety of the diverse assay types (Fig 2) generated as a part of the different ENCODE awards. The main assay types include: DNA binding assays, transcription assays, DNA accessibility assays, single cell assays, 3D chromatin structure assays, RNA binding assays, DNA methylation assays and assays that decipher the RNA structure. A full breakdown of all the genomics assays that belong to each of these assay types can be found in Fig 2.

Most ENCODE awards conducted experiments on humans and mice, with a few exceptions that were performed in cell-free or *in vitro* systems (Fig 1B). The experiments from modERN and modENCODE projects have data from three different Drosophila species (with majority being from *Drosophila melanogaster,* and a handful from two other species) as well as from the worm *Caenorhabditis elegans.* In addition, these assays have been performed on different biosample types including tissues, cell lines, primary cells, *in vitro* differentiated cells and organoids (Fig 1C).

*[Note: The landing page for the Experiment search and report views hide control assays and assays performed on perturbed biosamples. Using the facets these experiments can be viewed or hidden using the "hide control experiments" and "Perturbation" options respectively. The latter facet helps filter assays that were performed on modified (treated or genetically altered) samples created for the purpose of testing the effects of the perturbation on the biosample].*

## Uniformly processed data:

In addition to storing raw fastq files and metadata that is relevant to each of the experiments, the experiment objects also contain processed data files. Experiments may either host only lab processed data files or only ENCODE pipeline based uniformly processed data files. In some experiments though, there could also be a combination of both types. The uniformly processed pipeline data files are available for a key set of assays that make up the bulk of the ENCODE experimental corpus including: Histone ChIP-seq, TF ChIP-seq, MINT ChIP-seq, ATAC-seq, DNase-seq, RNA-seq, micro RNA-seq, long read RNA-seq, HiC and WGBS assays (https://www.encodeproject.org/data-analysis/). The software implementation of uniform processing pipelines was done as a collaborative effort within the ENCODE consortium and the code for all the pipelines is freely available for public use on github at https://github.com/ENCODE-DCC. More details on the uniform processing pipelines can be found in a separate publication[4]. In addition to displaying (for every dataset available on the portal) the relation between the various processed files generated by an execution of a pipeline run, the file association graphs also display useful quality metrics and summary statistics[5].

## Functional Characterization Experiments:

The Functional Characterization experiments include various assays that were performed as validations of the activity of predicted functional DNA elements (https://www.encodeproject.org/functional-characterization-matrix/?type=FunctionalCharacterizationExperiment&type=FunctionalCharacterizationSeries&type=TransgenicEnhancerExperiment&config=FunctionalCharacterization&datapoint=false&control_type!=*&status=released). Many of these predicted DNA elements were selected for analysis by utilising the processed data outputs of functional genomics based assays. These assays can be of two types: A) sequencing based methods and B) imaging based methods.

A total of 711 sequencing based datasets (including relevant controls) exist on the ENCODE portal that are classified as the Functional Characterization Experiments. CRISPR screens are grouped into Functional Characterization Series to demonstrate the relation between several different readouts. As shown in Fig 3, the various assays performed under this umbrella include several flavours of CRISPR screens[10] such as Flow-FISH CRISPR screen, proliferation CRISPR screen and FACS CRISPR screen as well as high-throughput Reporter Assays such as MPRA[11] and STARR-seq[12]. This dataset type also includes single cell assays (such as single nuclear ATAC-seq or snATAC-seq and single cell RNA-seq or scRNA-seq) performed in cell lines perturbed using CRISPR machinery, sometimes also referred to as Perturb-seq[13] or SPEAR-ATAC[14]. A majority of the assays that belong to this category were conducted in human biosamples (cell lines, primary cells, *in vitro* differentiated cells and organoids) along with a few from mouse cell lines (Fig 3).

There are 311 imaging based transgenic enhancer reporter experiments (https://www.encodeproject.org/search/?type=TransgenicEnhancerExperiment&status=released) performed in mice embryos (E11.5 days). The construct tested in these datasets consists of an enhancer element cloned upstream of a promoter sequence and the element activity is recorded using a beta-galactosidase reporter[15]. All the microscopic image outputs are provided by the ENCODE portal as attached characterization objects on the respective datasets and can be viewed by clicking on the image. The data can also be viewed on the Vista enhancer browser[16] (http://enhancer.lbl.gov/).

## Single cell datasets:

Single cell based genomics assays can be found on the Single Cell page (https://www.encodeproject.org/single-cell/?type=Experiment&assay_slims=Single+cell&status=released). This page is further divided into three different tabs to separately support high-throughput, perturbed high-throughput and low-throughput assays. The high-throughput tab is further divided into human and mouse with interactive body maps. Both human and mouse tabs contain datasets for snATAC-seq and scRNA-seq. In addition to these two assay types, the mouse tab also has long read scRNA-seq. The Perturbed high-throughput tab contains links to the perturb-seq-like functional characterization experiments mentioned above. The low-throughput tab contains single cell series that were performed using the Fluidigm C1 System.

## Series datasets:

A Series is a grouping of datasets that share a biological or functional theme. A dedicated landing page for the most popular series was developed to help explore them easily, and is subdivided into seven different tabs. The link for this page can be found here: https://www.encodeproject.org/series-search/?type=OrganismDevelopmentSeries. The seven different tabs include: organism development, treatment time, treatment concentration, replication timing, gene silencing, differentiation and disease. Other notable series that are not displayed on this page are the reference epigenomes series and multiOmic series.

## Organism Development series:

The organism development series were developed to display related functional genomics datasets performed to study the changes that occur within the genomic landscape of a particular biosample type during developmental stages of an organism. A total of 433 series of this type are found on the ENCODE portal.

Most of these series (total 401) come from experiments conducted on mouse tissues collected at various time-points of either embryonic, postnatal or adult developmental stages. For example: The dataset ENCSR275HXV (https://www.encodeproject.org/organism-development-series/ENCSR275HXV/) consists of 10 Histone ChIP-seq experiments (target: H3K27ac) and their corresponding control experiments that were performed on *M. musculus* heart tissue collected from strains, B6NTac;B6NCrl/Lap from 7 different embryonic time-points (10.5, 11.5, 12.5,13.5, 14.5, 15.5, 16.5 days) as well as one post-natal (0 days) and one adult (2 months) time point. The mouse organism development series have also been organised as a separate collection page, called the mouse development matrix. It summarises the 401 series and underlying 1934 functional genomics assays on a single matrix view (more details below).

The remainder of the development series consist of 35 *C. elegans* and 4 *H. sapiens* datasets. All of the *C. elegans* datasets were performed as a part of the modERN project and includes RNA-seq based assays performed at different developmental time points. With an exception of three series (ENCSR968SQN, ENCSR085DYI and ENCSR235YWK) that were assayed using a wild type isolate, the rest of the series contain results from assays that were performed on mutagenized strains repressing specific genes of interest. For example, ENCSR871ZUL consist of 6 RNA-seq experiments performed on a strain that repressed the homeobox protein, dve-1, and sampled at different larval stages including: 50, 80, 490, 520, 650 and 680 mins.

## Treatment time and Treatment concentration series:

The treatment time series organise functional genomics assays that were performed on the same biosample but sampled at different treatment durations (https://www.encodeproject.org/series-search/?type=TreatmentTimeSeries&status=relea

sed). There are a total of 82 datasets from treatment time series currently available on the ENCODE portal, comprising a total of 24 different treatments on different biosamples. The treated biosamples used in these assays include cell lines and *in vitro* differentiated cells from *H. sapiens*, primary cells from both *H. sapiens and M. musculus* and a tissue sample from *M. musculus*. The highest number of human data based series consist of datasets sampling 17 different chemical treatments on K562 cell lines, sampled at different time points. The second most assayed cell line includes A549 sampled at several time points of dexamethasone treatment. Example: https://www.encodeproject.org/treatment-time-series/ENCSR210PYP/ series lists all the ChIP-seq datasets with the target NR3C1 (nuclear receptor subfamily 3 group C member 1) performed on A549 cell lines with 100 mM of Dexamethasone for 0.5, 1, 2, 3, 4, 5, 6, 7, 8, 10, 12 hours and a non-treated control. Each of these datasets' corresponding input library controls are also listed in the same series.

The treatment concentration series (https://www.encodeproject.org/series-search/?type=TreatmentConcentrationSeries&status=released), on the other hand, organises datasets that were performed on a particular biosample for a fixed time but treated with different concentrations of the same chemical. Currently there are only 2 series available on the ENCODE portal for this series type. Example: https://www.encodeproject.org/treatment-concentration-series/ENCSR989EXF/ is a series comprising of 5 ChIP-seq datasets performed on A549 cell lines that were either treated with 0.5, 5 and 50 nM of dexamethasone for 1 hour (Treatment experiment) or with 0.02% ethanol (Treatment control). The 3 dexamethasone treated assays along with one ethanol treated assay within this series have a target gene NR3C1.

## Replication timing series:

The 22 datasets belonging to the replication timing series (https://www.encodeproject.org/series-search/?type=ReplicationTimingSeries&status=released) consist of datasets performed on 22 different primary cells and cell lines sampled at cell cycle phases including: early S, late S, G1b, S1, S2, S3, S4 and G2. The 20 biosamples from *H. sapiens* assayed in these experiments include 18 cell lines such as: A549, BG02, BJ, Caki2, GM06990, GM12801, GM12812, GM12878, G401, HeLa-S3, IMR-90, K562, LNCAP, MCF-7, NCI-H460, SK-N-MC, SK-N-SH, T47D as well as primary cells endothelial cells of umbilical vein and keratinocyte. The remaining two series come from *D. melanogaster* cell lines S2 and ML-DmBG3-c2.

## Gene silencing series:

In general, the 1,196 gene silencing series (https://www.encodeproject.org/series-search/?type=GeneSilencingSeries&status=relea

sed) are of two types: A) "silencing" using siRNA, shRNA, or CRISPR editing and B) depleting a particular protein using an auxin inducible degron (AID) method[17,18].

For the first category (A), each individual series consists: 1) an experimental assay wherein a specific gene of interest is knocked out by either CRISPR based genome editing technology or an RNAi based gene silencing technique, and 2) one or more equivalent wild type control assays performed on a non-modified cell line. About 90% of these are knockdowns or knockouts of RNA binding proteins followed by RNA-seq in both HepG2 and K562 and the remainder 10% are knockouts of transcription factors followed by RNA-seq (in K562, with one factor ZFX knocked out in C4-2B and MCF-7 cell lines). Several of these assays were conducted as part of the ENCORE project[19] and can also be found on the ENCORE matrix (more details below).

In the second category (B), each series is a combination of an experimental assay wherein auxin based induction was used to initiate protein degradation along with an untreated control assay in HCT-116 cells. There are 11 different assays performed by targeting 8 different proteins including CTCF, BRD4, CDK7, RAD21, SMARCA5, MED14, POLR2A and SUPT16H. These series are also displayed on a special collection page, known as the Protein knockdown or the Degron matrix (more details below).

### Differentiation series:

The differentiation series consists of experiments wherein *in vitro* differentiated cells or organoids were derived from embryonic stem cells (such as H9) or pluripotent stem cells (such GM23338 and WTC11). Currently there are 41 differentiation series on the ENCODE portal (https://www.encodeproject.org/series-search/?type=DifferentiationSeries&status=released). The seven main assay categories found within datasets for these series include MINT-ChIP-seq, ChIP-seq, polyA-plus RNA-seq, DNase-seq, ATAC-seq, RNA-seq and STARR-seq. The biosamples assayed within these series include human cell lines such as H9, GM23338, WTC11 and the derived *in vitro* differentiated cells including nephron progenitor and neural progenitor cells, as well as brain and nephron organoids sampled at different organoid growth time points. Example: https://www.encodeproject.org/differentiation-series/ENCSR015LDD/ series consists of 5 H3K4me3 experiments (from a MINT-ChIP assay), with one of the experiments performed on H9 stem cells, another assay on H9 derived nephron progenitor cells (8 days post differentiation) as well as 3 datasets from H9 derived nephron organoids sampled at 3 different time points (21, 35 and 49 days post differentiation). Another series https://www.encodeproject.org/differentiation-series/ENCSR318BKN/ consist of WTC11 cell lines at different stages of cardiomyocyte differentiation. In addition, there are 9 mouse differentiation series that consist of several immune cells sampled before and after being exposed to inflammation.

## Disease series:

The vast majority of ENCODE human samples are derived from healthy or putatively healthy donors. The exceptions include samples that comprise the 132 disease series (https://www.encodeproject.org/series-search/?type=DiseaseSeries&status=released). Most of the datasets belonging to these series are assays performed on brain samples from the Rush Institute collected from patients who had been diagnosed with different levels of cognitive impairment. These include: mild cognitive impairment, cognitive impairment and Alzheimer's disease. All of the cognitive impairments containing series consist of data that comes from an individual patient and can contain more than one assay type. For example: https://www.encodeproject.org/disease-series/ENCSR972RJG/ series consists of several assays including ChIP-seq, RNA-seq, DNase-seq performed on two different brain regions: head of caudate nucleus tissue and dorsolateral prefrontal cortex contributed by the same donor. In addition, there are 12 series from samples collected from patients with basal cell carcinoma and squamous cell carcinoma. Six different assays were performed on these different diseased samples including: RNA-seq, DNase-seq, ChIP-seq, microRNA-seq, long read RNA-seq and HiC. These series contain assays performed on two samples from the same donor: a tumour sample, and a matching healthy sample. Example: https://www.encodeproject.org/disease-series/ENCSR832CNS/.

## Reference epigenome series:

The reference epigenomes series (https://www.encodeproject.org/search/?type=ReferenceEpigenome&status=released) consists of experiments from ENCODE3 and the Roadmap for Epigenomics Consortium (REMC)[20]. A reference epigenome is defined as a collection of datasets profiling all the epigenomic marks on a particular biosample, ideally collected from the same individual donor. With 233 series originating from human biosamples and 92 from mouse biosamples, these 325 reference epigenomes make up a very useful collection. All the underlying 4,146 datasets are also presented in a matrix format on the Reference epigenomes matrix page (see more details below).

The human reference epigenome series contains assays performed on 109 tissues, 54 cell lines, 51 primary cells and 24 *in vitro* differentiated cells, while the mouse data comes from 86 tissues, 4 cell lines and 2 primary cells. The mouse data is available for assays such as Histone ChIP-seq (performed on 8 histone marks including 6 core marks: H3K4me1, H3K4me3, H3K27ac, H3K27me3, H3K36me3, H3K9me3), polyA plus RNA-seq, WGBS, microRNA-seq, ATAC-seq, TF-ChIP (4 core targets including CTCF, POLR2A, EP300 and POLR2AphosphoS5), DNase-seq and total RNA-seq. The ENCODE3 (non-roadmap) human reference epigenomes consists of assays such as: Histone ChIP-seq (12 marks including 6 core marks), TF ChIP-seq (4 core targets), DNase seq, total RNA-seq, ATAC-seq, small RNA-seq, microRNA-seq, WGBS, polyA plus RNA-seq, RRBS and microRNA counts. The Roadmap-based reference epigenome datasets contain Histone ChIP-seq data profiling 31 different histone marks (including the above listed 6 core marks) in addition to poly A plus RNA-seq, DNase-seq, RRBS, WGBS, RNA microarray, microRNA-seq, MRE-seq, MeDIP-seq and small RNA-seq . Example: https://www.encodeproject.org/reference-epigenomes/ENCSR970ENS/ contains 31 datasets profiling the epigenome of common myeloid progenitor, CD34-positive cells sampled from composite male and female human donors at different ages.

## Multiomics series:

The multiomics series contain 151 series that summarise the paired multiomics assays performed during the ENCODE4 phase on 139 tissues along with a couple cell lines, primary cells and *in vitro* differentiated cells obtained from both human and mouse samples (https://www.encodeproject.org/search/?type=MultiomicsSeries&status=released). Most series consist of paired scRNA-seq and snATAC-seq assays, with an exception of 3 assays from mice that contain pairs of long read RNA-seq and snATAC-seq. Example: https://www.encodeproject.org/multiomics-series/ENCSR032YYX/ series contains data from paired long read and short read Split-seq experiments performed on nuclear fraction of *in vitro* differentiated myotube cells.

## Computational and integrative products (Annotations):

With 67,008 annotation datasets currently on the portal, these make up the largest number of datasets found on the ENCODE website (https://www.encodeproject.org/search/?type=Annotation&status=released&award.project!=community). Annotation datasets are home to data generated in different ENCYCLOPAEDIA[21] versions including current, v4, v3, v2, v1, v0.3 and v0.2 as well as several other computationally predicted genome annotations derived using integrative analysis of several thousands of processed outputs files from functional genomics assays. A large proportion of the community datasets submitted to the portal host files that are utilised in a sister project called RegulomeDB[22] that has calculated the functional significance of SNPs (https://regulomedb.org/regulome-search/)

A special encyclopaedia browser was developed to visualise the various annotations on a genome browser along with the ability to select specific tracks of interest

([https://www.encodeproject.org/encyclopedia/?type=File&annotation_type=candidate+Cis-Regulatory+Elements&assembly=GRCh38&file_format=bigBed&file_format=bigWig&encyclopedia_version=current](https://www.encodeproject.org/encyclopedia/?type=File&annotation_type=candidate+Cis-Regulatory+Elements&assembly=GRCh38&file_format=bigBed&file_format=bigWig&encyclopedia_version=current)). The landing page is divided into two tabs - one tab for humans and the other for mouse, both of which display the various encyclopaedia annotations from different biosamples. A body map based filtering is available on both tabs and allows one to filter the tracks for specific organs or biosamples of interest. In addition, the tracks can also be filtered down using the biochemical activity facet to display only one (or any) of CTCF, DNase-seq, H3K27ac, H3K4me3 derived annotations. The browser tracks can also be sorted either by biosample term name or annotation type using the sort buttons. Navigation to specific gene locations can be performed using the search by gene box. An ability to navigate to specific chromosome locations (using the genome coordinates) is also available on the top most section of the browser. The cCRE tracks are coloured to display the predicted biochemical activity and the legends of each can be found on the top right corner of the genome browser.

# ENCODE Collections:

Collection pages are dedicated interfaces that display specific sets of functional genomic experiments. We have a total of 11 collection pages which are described below (Fig 5). Based on data obtained using google analytics, between 1st Jan 2022 and 31st Dec 2022, the five top-most visited collection pages include the ENCORE matrix, ENTEx matrix, Rush Alzheimer's Disease Study matrix, Mouse Development matrix and Deeply Profiled Cell Line matrix.

### ENCORE matrix:
The ENCORE matrix page ([https://www.encodeproject.org/encore-matrix/?type=Experiment&status=released&internal_tags=ENCORE](https://www.encodeproject.org/encore-matrix/?type=Experiment&status=released&internal_tags=ENCORE)) displays 1,832 datasets that were performed as a part of the ENCORE project[19]. ENCORE studies the binding of specific proteins to RNA using different assays such as shRNA RNA-seq, CRISPR RNA-seq, eCLIP and iCLIP. There are also some ChIP-seq and RNA-seq experiments. The matrix view shows all the gene targets along the y-axis and the cell line and assay type along the x-axis. Experiments performed in K652 are shown in blue, while experiments performed in HepG2 are shown in yellow. RBNS experiments are performed in a cell free system, and are represented as grey blocks on the matrix. Since the list of targets is long, this page comes with a filter box which can be used to narrow down the matrix. Example, typing CTCF in the filter box narrows down the matrix to just display the rows with CTCF as a target. This matrix also contains polyA-plus RNA-seq and total RNA-seq assays that were performed on RNA isolated from different subcellular compartments; these are shown as a separate matrix on the right.

**Immune cells:**

The immune cells matrix page summarises more than 1,736 datasets performed on primary immune cells from the ENCODE, Roadmap, and GGR projects ([https://www.encodeproject.org/immune-cells/?type=Experiment&replicates.library.biosample.donor.organism.scientific_name=Homo+sapiens&biosample_ontology.cell_slims=hematopoietic+cell&biosample_ontology.classification=primary+cell&control_type!=*&status!=replaced&status!=revoked&status!=archived&biosample_ontology.system_slims=immune+system&biosample_ontology.system_slims=circulatory+system&config=immune](https://www.encodeproject.org/immune-cells/?type=Experiment&replicates.library.biosample.donor.organism.scientific_name=Homo+sapiens&biosample_ontology.cell_slims=hematopoietic+cell&biosample_ontology.classification=primary+cell&control_type!=*&status!=replaced&status!=revoked&status!=archived&biosample_ontology.system_slims=immune+system&biosample_ontology.system_slims=circulatory+system&config=immune)). The matrix is subdivided into five major immune cell types along the y-axis: T cells (37 different cell types), B cells (7 different cell types), myeloid cells (8 different cell types), NK cells (2 different cell types) and mononuclear cells (2 different cell types). The x-axis represents the various assay types including: different flavours of RNA-seq, snATAC-seq, ATAC-seq, DNase-seq, various flavours of ChIP-seq (TF-ChIP, Histone ChIP and MINT-ChIP), ChIA-PET, WGBS, intact HiC, Repli-Chip. The ChIP-seq and ChIA-PET assays are further broken down into different columns based on their targets.

The upper left section of the page allows the matrix to be filtered by specific cell lineages and helps show the relationship between immune cell subtypes. Specific sections of this cell lineage graph can be hidden or expanded using the (-) or (+) buttons present on the higher level nodes. Clicking on an individual cell drawing image, allows one to either select or de-select that cell type. There are also "Show all" and "Hide all" buttons to facilitate the filtering process. For example, if interested in only two specific cell types, one can select the "Hide all" button and then select those two specific cells. Since some cells have been sampled from patients with Multiple Sclerosis, below the expandable cell-lineage modal, we provide a filter that allows them to be included or excluded.

**Stem cells development matrix:**

The Stem cells development matrix contains four tabs for different stem cells along with their *in vitro* derived cell types and organoids ([https://www.encodeproject.org/stem-cell-matrix/?type=Experiment&replicates.library.biosample.donor.accession=ENCDO222AAA&status=released&control_type!=*](https://www.encodeproject.org/stem-cell-matrix/?type=Experiment&replicates.library.biosample.donor.accession=ENCDO222AAA&status=released&control_type!=*)). The four tabs include: a) H1 derived human embryonic stem cells (hESCs), b) H9 derived hESCs, c) PGP donor (GM23338) based induced pluripotent stem cells (iPSCs) and d) H9 samples that were collected as a part of the Southeast Stem Cell Consortium (SESCC) effort. [*Note: The SESCC tab contains a subset of the experiments available in the H9 tab*]. Currently, the total number of assays available for both H1 and H9 matrices is 313, 156 for the PGP matrix and 81 for the SESCC matrix. The H9 and PGP tabs contain data from the base cell lines, *in vitro differentiated* cells as well as organoids, while the H1 and SESCC tabs contain data only from the former two biosample types. With 39 different biosamples assayed using 30 functional genomics assays, this collection serves as a very useful resource for studying epigenetic changes that transpire during *in vitro* differentiation of these classical cell lines.

Just like the immune cells matrix page, the stem cells matrix page includes both 1) a matrix of various cell types along the x-axis and the various assays along the y-axis as well as 2) a cell lineage node graph that displays the relationships between the progenitor stem cell and all the cell types that were derived from it. The cells are colour coded to indicate the different developmental lineages (such as Ectoderm, Mesoderm and Endoderm). Similar to the Immune cells matrix page, an option to expand or collapse the nodes using the (+) or (-) buttons is available above the cell drawings as well as an ability to hide/show rows from the matrix by selecting/deselecting cell node drawing. Again, like the other matrices described above, the rows, columns and individual cells in the matrix are clickable for further data exploration on the experiments search page.

## Deeply profiled cell lines:

The deeply profiled cell lines page represents data from 16 cell lines that have been extensively studied by ENCODE. The page is divided into two subviews of a matrix with a radio button that allows a switch between "All" (5501 datasets) and "Uniform batch growth" (190 datasets). The Uniform batch growth view displays a set of experiments performed on cells grown as a uniform batch and distributed to different labs for different assays. Such a set-up to share cell-growths was an attempt to reduce batch effects that may arise from growing cells in different labs (McShane et al., "Integrating nascent transcription and chromatin conformation across human cell lines", in preparation).There is a special emphasis on dynamic RNA measurements including PRO-seq, PRO-cap, Bru-seq, BruUV-seq and BruChase-seq. The cell lines that are a part of the Uniform batch growth are : A673, Caco-2, Calu-3, GM12878, HCT116, HepG2, IMR-90, K562, MCF10A, MCF-7, OCI-LY7, PC-3, PC-9, Panc1 as well as two primary cells: endothelial cells of umbilical vein (HUVEC) and mammary epithelial cells.These different cell types are displayed on the matrix along the y-axis and the various assays are displayed along the x-axis. The uniform growth batch identifiers are listed as sub-categories below the corresponding cell type. Intact HiC, ATAC-seq, DNase-seq, ChIA-PET, long read RNA-seq, micro RNA-seq, total RNA-seq and snATAC-seq were also performed on these growths, in addition to the dynamic RNA experiments mentioned above. In addition to the Uniform batch growth matrix, the "All" tab contains data on these 16 cell lines from all the phases of ENCODE.

## Protein Knockdown/Degron matrix:

There are 252 datasets that are summarised on the Degron matrix (https://www.encodeproject.org/degron-matrix/?type=Experiment&control_type!=*&status=released&internal_tags=Degron). This matrix displays experiments that were performed on transformed HCT116 cell lines, utilising the auxin inducible Degron (AID) system[23]. The chromatin modulating proteins targeted for the knockdown assays include CTCF, BRD4, CDK7, RAD21, POLR2A, SMARCA5, MED14 and SUPT16H. The assays performed on these transformed lines include: MINT-ChIP-seq (H3K27ac, H3K4me1, H3K4me3, H3K9me3, H3K27me3 and H3K36me3), ChIA-PET (CTCF and POLR2A), ATAC-seq, Bru-seq, PRO-seq, TF-ChIP-seq (CTCF), DNase-seq, PRO-cap and snATAC-seq. Most of the assays have been performed in pairs: one of the pairs is a non-treated control experiment while the other was treated with 1 µM 5-Phenyl-1H-indole-3-acetic acid for 6 hours to induce the AID system. The presence or absence of data is indicated by coloured blue cells in the matrix, with a lighter shade indicating the presence of only one paired experiment. This matrix also includes a search box on the top to aid with filtering of datasets using search terms.

## Human donor matrix:

The Human donor matrix provides a matrix view in which individual rows represent a single human donor along the y-axis and the functional genomics experiments performed on samples from this donor along the x-axis (https://www.encodeproject.org/human-donor-matrix/?type=Experiment&control_type!=*&replicates.library.biosample.donor.organism.scientific_name=Homo+sapiens&biosample_ontology.classification=tissue&status=released&config=HumanDonorMatrix). With a total of 6944 datasets, this matrix provides a view to identify comparable datasets that were assayed using biosamples collected from the same individual donor. Each row on the matrix starts with a donor accession (starting with an "ENCDO" accession id) followed by their age, sex and disease information (if any reported). The coloured cells per row on the matrix indicates the total number of experiments performed for each assay type on a biosample collected from the donor represented on the x-axis. The matrix by default is limited to show tissues, with an ability to switch to primary cells or both using the Biosample classification toggle.

## Epigenomes from four individuals or ENTEx matrix:

The ENTEx matrix (https://www.encodeproject.org/entex-matrix/?type=Experiment&status=released&internal_tags=ENTEx) shows samples collected as a part of a collaborative effort with the GTEx consortium[24]. The matrix shows the full spectrum of assays performed for tissue samples that are derived from 4 human donors. The donors included in this matrix are two adult males with donor ids ENCDO845WKR (age: 37 years) and ENCDO451RUA (age: 54 years) as well as two adult females with donor ids ENCDO793LXB (age: 53 years) and ENCDO271OUW (age: 51 years). The presence of an assay performed on a particular biosample from one of the four donors is represented by a different coloured slice on a 4-way pie-chart. The ENTEx matrix consists of 1567 experiments performed on 31 unique tissue samples. The different assay types available for this collection includes flavours of RNA-seq, RAMPAGE, microRNA counts, Histone ChIP-seq (6 histone marks), TF ChIP-seq (4 targets), eCLIP (2 targets), ATAC-seq, snATAC-seq, DNase-seq, WGBS, DNAme array, RAMPAGE, genotyping array, WGS and in situ HiC.

## Rush Alzheimer's Disease Study Matrix:

The Rush Alzheimer's disease (AD) study matrix (https://www.encodeproject.org/brain-matrix/?type=Experiment&status=released&internal_tags=RushAD) consists of a set of experiments performed on three brain tissue sections collected from donors with neurological disorders, alongside control (non-diseased) brains. The three brain tissues, shown as three different colours on the pie chart includes Dorsolateral prefrontal cortex (DLPFC), head of caudate nucleus and the posterior cingulate gyrus. Similar to the donor-centric matrix, each row comprises of unique donor accession which starts with an "ENCDO" accession ID, along with donor sex and age at the time of sampling.

The matrix is further divided into 5 sections based on the type of the neurological disease. These sections include: No cognitive impairment (donors without any known impairment), Mild Cognitive, Cognitive, Alzheimer's disease + Cognitive as well as Alzheimer's disease donors. The presence or absence of different tissue sections from each donor is evident by the coloured pie-charts. Similar to the other matrices, clicking on individual rows redirects to the experiment search page for that donor. Clicking on the pie-charts shows search results for the combination of the assay and that donor. The various assays that were performed on the Rush Brain samples include: DNase-seq, Histone ChIP-seq (H3K27ac, H3K4me3 and H3K27me3), total RNA-seq, TF-ChIP-seq (CTCF), micro RNA-seq, long read RNA-seq and intact HiC. Except for the DNase-seq, the majority of the other assays were performed on the DLPFC.

## Reference Epigenomes matrix:

The reference epigenomes page displays a matrix view of all the reference epigenomes (https://www.encodeproject.org/reference-epigenome-matrix/?type=Experiment&control_type!=*&related_series.@type=ReferenceEpigenome&replicates.library.biosample.donor.organism.scientific_name=Homo+sapiens&status=released). The page is subdivided into two matrices: human and mouse, each having an interactive body map image with the ability to sub-select organs or organ-systems of choice. As described above since the human reference epigenomes comprise of data from Roadmap and ENCODE, the matrix comes with a radial button to switch between displaying "All", "Roadmap" or "Non-Roadmap" datasets. Both the human and mouse matrices are further subdivided into tissues, cell lines, primary cells and *in vitro* differentiated cells along the x-axis. Since the list of each type is long, only the top 10 entries are displayed by default for each sub-category and these can be expanded for further viewing. Horizontal scrolling along the x-axis allows visualisation of the long list of assays while having the biosamples fixed along the y-axis.

**Mouse development matrix:**

The mouse development matrix (https://www.encodeproject.org/mouse-development-matrix/?type=Experiment&status=released&related_series.@type=OrganismDevelopmentSeries&replicates.library.biosample.organism.scientific_name=Mus+musculus) summarises a rich collection of 1934 datasets from 39 different mice tissue samples, collected at different time points from three developmental stages- embryonic (17 time points), postnatal (6 time points) and adult (13 time points). Like most of the matrices, the y-axis represents the biosamples while the x-axis represents the different assays. However, the mouse development matrix has an additional dimension showing developmental stages and time-point based information, displayed as additional rows (shaded in 3 levels of yellow to represent each developmental stage) beneath the header row (shaded dark blue). The individual cells within the sub-section rows are shaded to indicate the presence (or absence) of data availability for that tissue + developmental stage + time point + assay combination.

# New home page

To provide an enhanced experience, we have completely remodelled the ENCODE Portal home page. The home page now provides a way to navigate through the wide variety of data types and collections available on the data portal. As demonstrated in Fig 6, the top section of the page contains a search box that can be used to find objects relating to the search input (see below for more details). Below the search box there are several clickable "cards" with a small logo, name, and description (which can be seen by selecting the question mark at the top left corner of the card). Upon selecting each of the individual cards, the respective landing pages for each can be viewed for further data exploration.

The cards on the homepage are laid out in four different panels. The top two panels (orange boxed cards in Fig 6) interact with the search box while the bottom two (green boxed cards in Fig 6) do not. The topmost panel consists of the three major ENCODE data-type cards: Functional genomics experiments, Functional characterization experiments and Encyclopaedia of elements (specifically representative DHSs, conserved cis-Regulatory Elements; cCREs, and chromatin state models). All the cards below the top-panel consist of useful subsets of the main data corpus organised on different pages (see above section on Collection pages). The functional genomics experiments card links to a matrix view displaying the various assays of this category organised by different biosamples. The functional characterization experiments card also links to a matrix view that displays all the assays classified by organism and biosamples. Similarly, the Encyclopaedia of elements card links to a search results page hosting the current version of the ENCODE encyclopaedia, including candidate *cis*-regulatory elements (cCREs), chromatin states and representative DNase hypersensitivity sites.

The cards located on the second panel include combinations of several ENCODE collections (more details above) such as Rush Alzheimer's disease study matrix, ENTEx matrix, Deeply profiled cell lines matrix, human donors matrix, ENCORE matrix and the stem cell development page. In addition, we have cards that link to the Computational integrative products matrix and a search page for Imputed datasets. The third panel consists of cards that link to other collections, series and applications that do not interact with the search box. The collections include Immune cells matrix, Mouse development matrix and Reference epigenomes matrix. The series cards include: Functional genomics series search and Single cell experiments search. In addition, we have the RNA-get, Region-search, Encyclopaedia browser and ChIP-seq experiments matrix. The fourth panel links to the various helpful pages such as Materials and methods, Publications and Getting-started.

## Search box usage:

The search box on the top-most part of the home page can be used to search either the ENCODE portal or the SCREEN[21] site (https://screen.wenglab.org). A toggle button is provided to select one of the two options. SCREEN is a tool that was developed to easily browse and search the candidate cis-Regulatory elements predicted by using the entire ENCODE data corpus.

For example, as shown in Fig 7, typing "H3K4me3" in the search box, and moving the toggle to ENCODE, the search for "H3K4me3" matches all cards that include any datasets containing this search term and are highlighted in yellow. In addition, a smaller box below the search-box will appear that displays a list of all the different object-types on the ENCODE portal that contains search results related to H3K4me3. Each of the items in the objects list is clickable and redirects to the corresponding page for that object type along with that search term appended in the query. Example, clicking on File will lead to https://www.encodeproject.org/search/?type=File&searchTerm=H3K4me3.

In contrast, when searching the SCREEN database, an option to select either human or mouse genes is shown. For example, searching for CTCF and clicking on the box "Human GRCh38" opens a new page that links to the following page: https://screen.wenglab.org/search/?q=CTCF&uuid=0&assembly=GRCh38. Once on this page, one can navigate to various tabs showing different results for CTCF from the SCREEN registry.

## Carts- a collection of user defined datasets:

Carts allow users to select custom combinations of datasets (including functional genomics, functional characterization, or computational annotations). A user can also add series (sets of datasets) to a cart, effectively adding all the member datasets from that series. Every dataset search page contains a cart icon next to the individual search results when using the search pages. Clicking on those cart icons adds the individual dataset to the cart one at a time. In addition, every search page contains a button on the top section which is named "Add all datasets to cart". Clicking on this button will let many datasets to be added to the cart at the same time. A single cart can contain a maximum total of 8000 datasets and each user is allowed 30 different carts at a given time. The total number of datasets added to a cart can be seen on the top menu bar under the cart icon. The carts feature is available only when a user is logged in. (Disclaimer: to login, users must register an email address with the ENCODE portal; there is no charge for using it and we do not share this email).

The different carts owned by a single user can be managed using the cart-manager page https://www.encodeproject.org/cart-manager/. This page lists all the existing carts and allows them to be annotated by adding a useful title and a short description. Each cart also has a unique identifier and a unique URL, hence, this cart URL references this custom collection and can be shared in publications or other documentation as needed. In addition, there is also an option of making your cart "Listed" which will allow a released cart to be viewed publicly on https://www.encodeproject.org/search/?type=Cart&status=listed&status=released.
There is a reversible "lock" to prevent it from being altered. Renaming and deleting of existing carts is also allowed from the cart manager. There are two main sections on the cart pages: Different tabs on the page allows one to view the files in various ways. Filters on the left hand side allow to further sub-select files of interest. The carts allow visualisation of genomic tracks from an arbitrary set of datasets on the Valis browser and to select subsets of files to download (rather than downloading by search result or individual dataset). To further explore all the cart features, refer to the video tutorials found listed here: https://www.encodeproject.org/help/cart/.

# RNA-get:

The RNA-get[25] page provides a way to obtain gene expression information for every human and mouse gene from polyA-plus-RNA-seq, polyA-minus-RNA-seq and total RNA-seq assays represented on the ENCODE portal (https://www.encodeproject.org/rnaget-report/?type=RNAExpression). The page has a tabular layout along with facets on the left. The facets help narrow down the results by biosample classification, biosample term name, donor organism, donor sex and the assembly used for generating these expression values. The report view is customizable using the button "columns" located at the top section of the page, columns can be selected and deselected based on the downstream data analysis. The Download tsv button allows a tab-separated file to be downloaded.

The default columns selected in the report table includes useful information about the genes, such as the Feature ids (ENSG gene ids), Gene names, Gene symbols, Gene titles as well as the expression values reported in terms of both: TPM (transcript per million) and FPKM (fragments per kilobase of exon per million mapped fragments) values. In addition, the report table also lists several useful metadata properties of the original assay whose files were used to generate the TPM/FPKM values for the gene listed in each row, such as the donor organism, biosample term name and corresponding biosample ontology based classification. Moreover, the source file and source dataset are listed using their ENCFF/ENCSR accession IDs and are hyperlinked for easy access.

**Glossary** (for more see: https://www.encodeproject.org/glossary/):

***ENCODE statuses*:** The status of an ENCODE object determines the visibility of the object and is also an indicator of the quality checks it has passed. Objects with "released", "archived" and "revoked" statuses are publicly available.
***Released*** datasets or files are the most recent and up-to-date versions of the objects. Released items have been carefully reviewed based upon the ENCODE consortium data quality standards. Released datasets processed once will have only one set (Analysis) of released files, while datasets processed multiple times (using different reference files or different versions of the pipeline) will have multiple sets (Analyses) of released files.
***Archived*** datasets or files indicate that the relevant object is valid and usable but the users will find a more up-to-date object on the portal equivalent to this.

***Revoked*** datasets or files have been released to the public at a previous time, but were later found to have some data quality or metadata issues. Revoked files or datasets should not be used but they remain publicly available for backwards compatibility.
For more information, readers can refer to our help pages located at
https://www.encodeproject.org/help/getting-started/status-terms/ and
https://www.encodeproject.org/data-standards/
*[Note: Even though statuses exist for every object type, here we will refer to only the dataset and file objects. In addition, several other statuses exist on the portal such as "in progress", "submitted", etc. However those are not publicly available and can only be viewed by admins].*

**Functional genomics experiments:** Classical genomics experiments that are performed on a biosample to study the transcriptional, epigenetic or regulatory state.
**Functional characterization experiments:** Experiments that are designed to study the effects of altering specific regions of interest of DNA. In many cases, these specific regions of interest are selected based on pre-existing knowledge that would have assigned these regions as being putative regulatory regions.
**Annotations:** Datasets representing computational predictions and integrative analysis outputs utilising functional genomics and functional characterization experimental results.
**Encyclopaedia:** A subset of integrative annotation datasets that were generated to define the ground level (open chromatin, histone mark enrichment, transcription factor binding, gene expression, promoter activity profiling, RNA binding occupancy, DNA methylation, 3D chromatin interactions and TADs) as well as integrative level annotations (such as registry of candidate cis regulatory elements and chromatin states). For more information, please refer to:
https://www.encodeproject.org/data/annotations/

**Targets:** A particular gene or protein that was the focus of a given assay. For more information on target categories and categorization:
https://www.encodeproject.org/target-categorization/ and
https://www.encodeproject.org/glossary/#target-categories

**ENCODE assays:**
**ChIP-seq**[26,27,28]**:** Chromatin Immunoprecipitation sequencing is an assay attempting to identify the genome-wide protein binding regions of specific proteins of interest by utilising antibodies to immunoprecipitate DNA-protein complexes followed by deep sequencing.
**TF ChIP-seq:** A ChIP-seq assay attempting to identify the genome-wide DNA binding regions of a particular transcription factor.

**Histone ChIP-seq:** A ChIP-seq assay attempting to identify the genome-wide DNA binding regions of specific histone marks.

**MINT ChIP-seq**[29]**:** A new multiplexed version of Histone ChIP-seq experiment developed by Bernstein lab. More details: https://www.encodeproject.org/documents/fc792518-31dc-4829-81d3-4dc9299c91bf/@@download/attachment/Bernstein-mint-chip3-a-low-input-chip-seq-protocol-using-multiplexed-chromatin-and-T7-amplificationn.pdf

**ATAC-seq**[30]: An assay profiling genome-wide chromatin accessible regions by utilising Tn5 transposase activity followed by tagmentation of open chromatin regions and DNA sequencing.

**DNase-seq**[31]**:** An assay profiling genome-wide open-chromatin regions by utilising DNaseI treatment to isolate protein bound followed by DNA sequencing.

**RNA-seq**[32]**:** Sequencing based assays attempting to quantify the abundance and sequences of all the existing RNA molecules within a biosample by first extracting intact RNA followed by cDNA conversion and NGS based sequencing.

**Total RNA-seq:** An RNA sequencing experiment attempting to quantify all the protein coding RNA molecules existing within a biosample.

**polyA-plus RNA-seq:** An RNA sequencing experiment that is performed on poly-A plus enriched RNA transcripts.

**polyA-minus-RNA-seq:** An RNA sequencing experiment that is performed on poly-A minus enriched RNA transcripts.

**small RNA-seq:** A variation of RNA sequencing technology attempting to quantify the abundance of small RNAs (size less than 200 nucleotides) present in a biosample.

**microRNA-seq:** A variation of RNA sequencing technology attempting to quantify the abundance of small non-coding regulatory RNAs (also known as microRNAs) present in a biosample.

**shRNA RNA-seq:** RNA-sequencing assay performed to study the effect of silencing a target protein of interest using RNAi technique.

**CRISPR RNA-seq:** RNA-sequencing assay performed to study the effect of knocking out a target protein of interest using CRISPRi based technology.

**PRO-seq**[33]**:** Precision nuclear run-on sequencing or PRO-seq is a sequencing assay that attempts to map RNA polymerase II pause sites.

**PRO-cap**[33]**:** A variation of PRO-seq that attempts to map the RNA polymerase II initiation sites.

**Bru-seq**[34]**:** Bromouridine sequencing or Bru-seq assay is a variation of RNA sequencing in which the nascent RNA transcripts are captured utilising bromouridine tagged UTP treatment prior to sequencing.

**BruUV-seq**[35]**:** BruUV-seq is a variation of Bru-seq that utilises UV radiation to aid with creation of random transcription blocking DNA lesions prior to bromouridine tagging treatment resulting in an enrichment of 5' nascent transcription start sites (TSS).

**BruChase-seq**[34]**:** Bromouridine Pulse-Chase sequencing or BruChase-seq assays attempt to quantify the relative stability of nascent RNA transcripts.
For more information refer to:
https://www.illumina.com/content/dam/illumina-marketing/documents/products/research_reviews/rna-sequencing-methods-review-web.pdf

**long read RNA-seq:** A long range (>10 kb) sequencing based technology attempting to quantify the abundance of RNA molecules within a given sample with an added advantage of having the ability to profile full length RNA molecules. For more information refer to:
https://longseq.cd-genomics.com/human-genomics-with-long-read-sequencing.html
and
https://www.encodeproject.org/documents/3baa46d2-cb88-4608-8877-70596d200489/@@download/attachment/ENCODE_longread_wetlab_protocolv3.pdf

**iCLIP**[36]**:** Individual Nucleotide Resolution CLIP (Cross linking and immunoprecipitation) is an assay that is used to identify protein-RNA interactions by immunoprecipitation of RNA-protein complexes followed by proteinase K treatment such that the residual RNA upon cDNA conversion gets truncated at the protein binding site.

**eCLIP**[37]**:** Enhanced crosslinking immunoprecipitation attempts to identify RBP (RNA binding proteins) binding sites on their target RNA molecules. This is a modified version of iCLIP and uses barcoded DNA adapters to reduce amplification bias.

**RBNS**[38]**:** RNA bind-n-seq assays attempt to characterise the sequence as well as structural specificity of RNA binding proteins.

**ChIA-PET**[39]**:** Chromatin interaction analysis with paired end tag sequencing attempts to identify genome-wide binding sites of transcription factors that bind DNA sequences over a long range chromatin region. This method uses Mme1 restriction enzyme to generate PET constructs followed by deep sequencing.

**HiC**[40]**:** A sequencing assay attempting to map the long range chromatin interactions existing within the nuclei of a given biosample by utilising formaldehyde to cross-link chromatin followed by solubilization, digestion, and re-ligation to produce chimeric DNA which is deeply sequenced to map long range chromatin interactions.

***In situ* HiC**[41]**:** An improved version of dilution HiC where the chromatin is cross-linked within intact nuclei leading to reduction in random ligation and overall improved signal. The primary method for chromatin fragmentation is the use of restriction enzymes.

**Intact HiC:** Most recent variant of Hi-C that maps fine-scale contacts like loops down to base-pair resolution and generates profiles of chromatin accessibility. Intact Hi-C achieves this by avoiding use of harsh reagents that can disrupt chromatin organisation and using nucleases that fragment DNA based on accessibility. For more information refer to:
https://www.encodeproject.org/documents/768fa33e-3c32-4ce2-8f78-b9fafda06cbc/@@download/attachment/intacthi-c_modularprotocol-release.pdf

**WGBS**[42,43]: Whole genome bisulfite sequencing OR WGBS utilises sodium bisulfite to treat genomic DNA which results in demethylation of unmethylated Cytosines within genomic DNA. The following deep sequencing allows detection of methylated cytosines in the genomic DNA by comparing bisulfite treated and untreated samples.

**RRBS**[44]: Reduced representation bisulfite sequencing or RRBS utilises one or multiple restriction enzymes based cleavage to enrich for genomic regions of interest followed by bisulfite sequencing to identify methylated sites within specific regions.

**MRE-seq**[45]: Methylation sensitive restriction enzyme sequencing or MRE-seq utilises different methylation sensitive restriction enzymes to digest genomic DNA followed by deep DNA sequencing to allow identification of methylation sites within the genome.

**MeDIP-seq**[46]: Methylated DNA immunoprecipitation sequencing or MeDIP-seq assay attempts to identify all the 5-methylcytosine and 5-hydroxymethylcytosine modifications by utilising antibodies targeting each followed by deep sequencing.

**WGS**[47]: Whole genome sequencing is a technique for sequencing entire genomes from a given biosample using short reads in order to identify variations between genomes .

**RAMPAGE**[48]: RNA Annotation and Mapping of Promoters for the Analysis of Gene Expression or RAMPAGE is a technique that aims to identify Transcription start sites (TSSs) along with promoter expression quantification and transcription connectivity by utilising highly multiplexed cDNA sequencing libraries.

For more information refer to this:
https://www.encodeproject.org/documents/3006c3aa-f1a5-4b16-9a78-c0a57aeb30ec/@@download/attachment/RAMPAGE%20pipeline%20overview.pdf

**microRNA counts:** A technique that relies on the nanostring nCounter digital analyzer to count the number of target miRNAs. For more details refer to:
https://www.encodeproject.org/documents/d90ef931-e735-4b25-a7f8-6ddbcaea71dd/@@download/attachment/NanoString%20miRNA%20Assay.pdf

**Repli-ChIP**[49]: An assay attempting to identify comparative differences between genome-wide changes that transpire during different replication timings of cells using 5-bromo-2-deoxyuridine (BrdU) followed by flow cytometry and collection of S-phase fractions. The purified BrdU-labelled DNA is then hybridised to a whole genome hybridization microarray to identify the comparative genome-level differences.

**genotyping array**[51]: Comparative genomic hybridization by array is an array based technique to identify single nucleotide polymorphisms (SNPs), copy number variations (CNVs), loss of heterozygosity (LOH) and insertion/deletions within a given biosample.

**RNA microarray:** An array-based assay that aims to detect actively expressed genes using microarray probes that are designed to bind unique sequences of interest.

**DNAme array**[50]: DNA methylation profiling by array is a genome-wide screening technique to identify CpG sites in the human methylome. For more information refer to:
https://www.illumina.com/content/dam/illumina/gcs/assembled-assets/marketing-literatur

e/infinium-methylation-epic-data-sheet-m-gl-01156/infinium-methylation-epic-data-sheet-m-gl-01156.pdf

**Single cell sequencing**[52]**:** Single cell sequencing is a technology that allows detection of genomic and transcriptomic variation at the single cell resolution. Unlike the bulk methods which provide an average signal over the entire biosample, the single cell sequencing technology provides high resolution variation information at every cell level.

**snATAC-seq**[53]**:** Single nucleus ATAC-seq combines single cell sequencing technology and ATAC-seq to provide information about open chromatin regions at single nuclear resolution.

**scRNA-seq**[54]**:** Single cell RNA-seq combines single cell sequencing technology and RNA-seq to provide transcriptome profiles at single cell resolution.

**long read scRNA-seq**[55]**:** Long read single cell RNA-seq combines long read technology along with scRNA-seq to provide transcriptome profiles utilising long read sequencing at single cell resolution.

**MPRA**[56]**:** Massively parallel reporter assays is a reporter assay that aims to interrogate the activity of several regulatory genomic regions simultaneously.

**STARR-seq**[12]**:** Self transcribing active regulatory regions sequencing attempts to quantify enhancer activity of millions of candidate DNA regions to create a genome-wide quantitative enhancer map.

*In vivo* **transgenic enhancer assays**[15]**:** *In vivo* transgenic enhancer assays are reporter assays that utilise colorimetric reporters to score the tested putative regulatory regions.

**CRISPR screens**[57]**:** CRISPR-Cas9 technology based genome scale screens attempting to identify the functions of specific regions of genomes by 1) introducing targeted genomic mutations using CRISPR technology leading to 2) either loss or gain of function of specific genomic regions followed by 3) testing the resulting effects under various selection pressures. For more information refer to:
https://www.synthego.com/guide/crispr-methods/crispr-screen

**Flow-FISH CRISPR screen**[10]**:** Flow-FISH (Fluorescent in situ hybridization) CRISPR screen is a variation of CRISPR screen which uses Flow cytometry as a selection method.

**Proliferation CRISPR screen**[10]**:** Proliferation CRISPR screen is a variation of CRISPR screen which uses cell proliferation as a selection method.

**FACS CRISPR screen**[10]**:** FACS CRISPR screen is a variation of CRISPR screen which uses FACS sorting as a selection method.

# Figures:

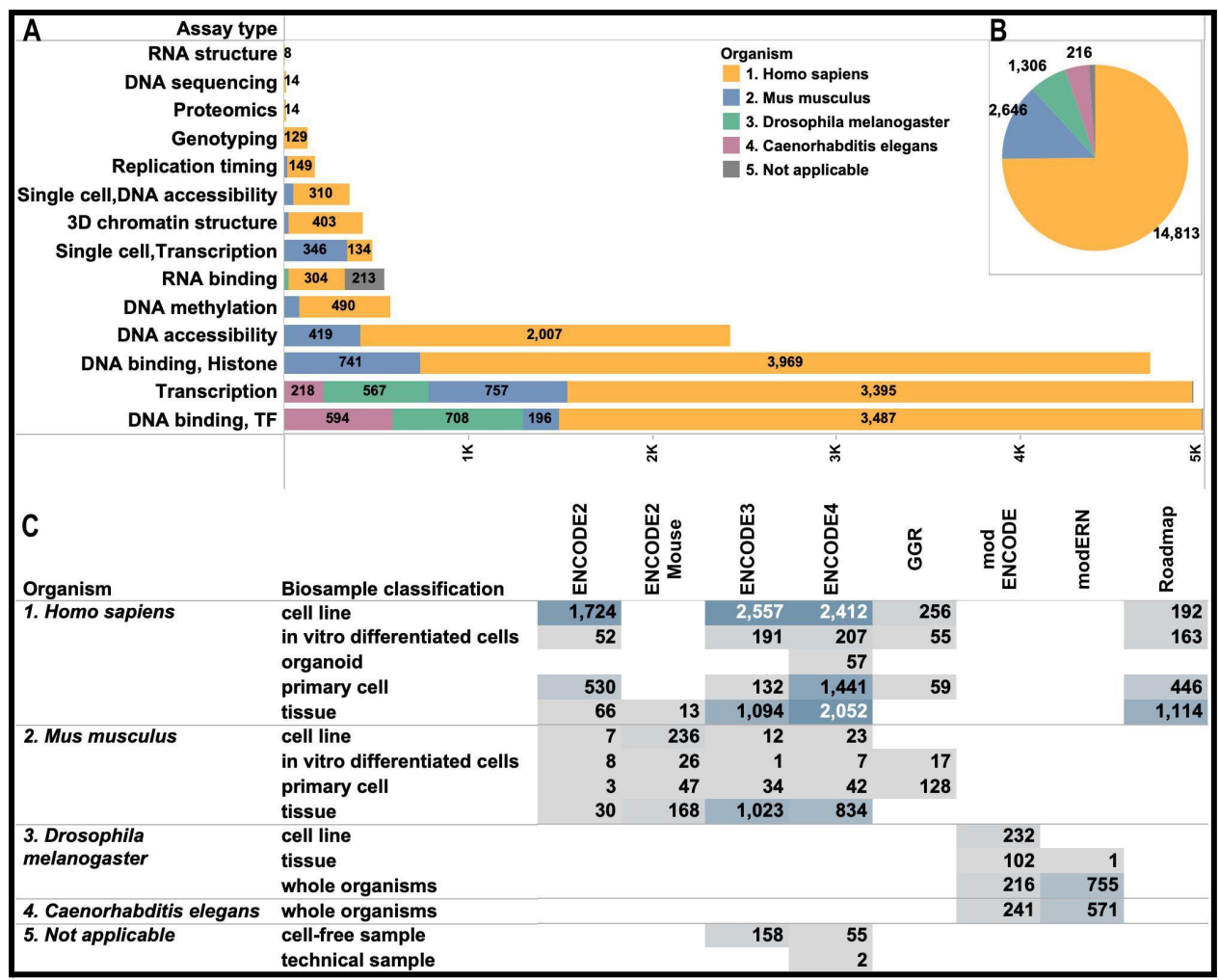

Fig 1: Plots showing diversity and the number of functional genomics experiments classified by A) Assay type B) Biosample organism C) Biosample classification and awards.

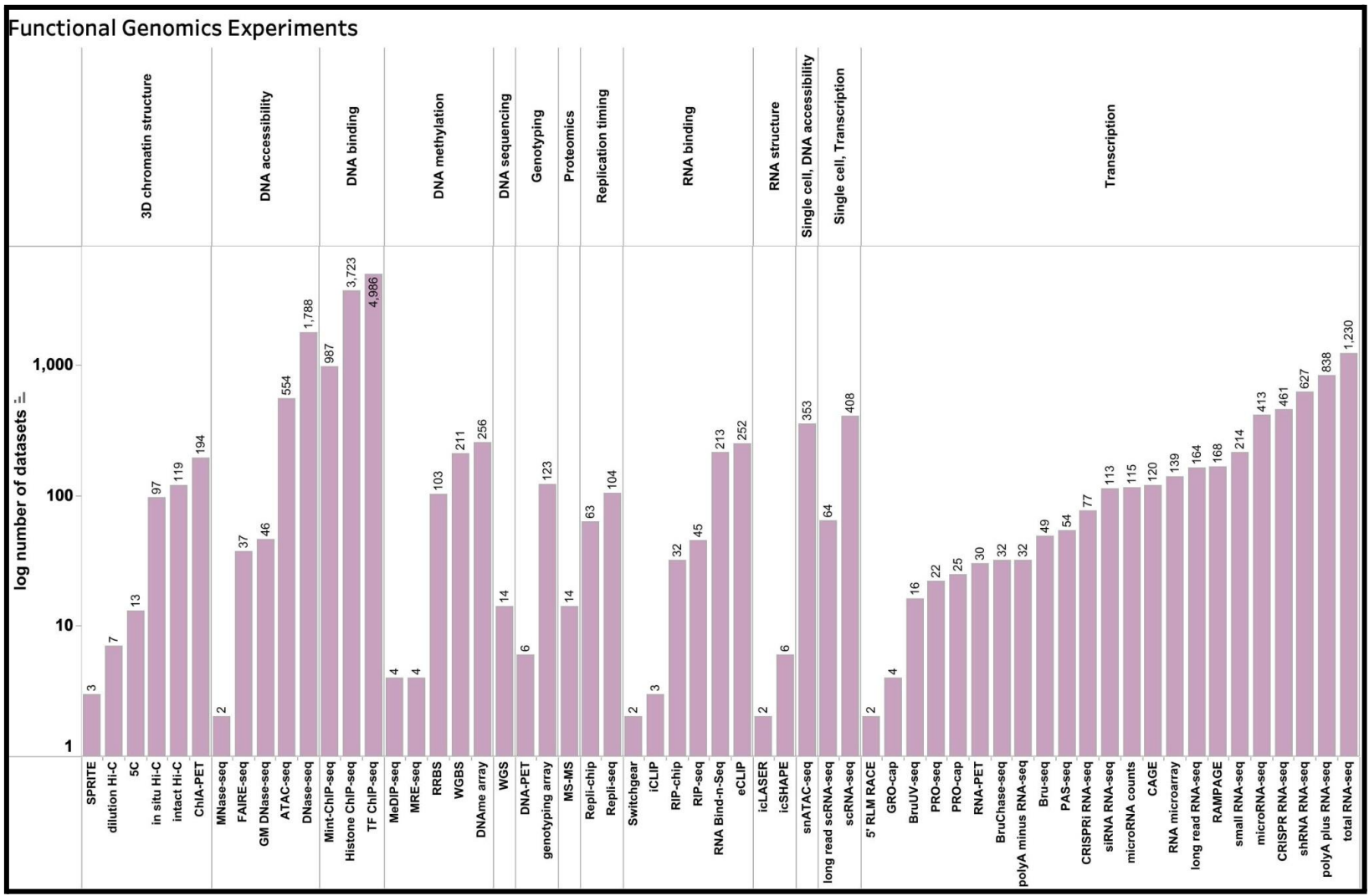

Fig 2: Plot displaying the number of diverse functional genomics experiments classified by different genomics assays

| Organism | Biosample Classification | Elements selection method | CRISPR screen | | | | Massively parallel reporter assay | | Single cell | |
| | | | CRISPR screen | FACS CRISPR screen | Flow-FISH CRISPR screen | proliferation CRISPR screen | MPRA | STARR-seq | perturbation followed by scRNA-seq | perturbation followed by snATAC-seq |
|---|---|---|---|---|---|---|---|---|---|---|
| *Homo sapiens* | cell line | accessible genome regions | | | | 6 | | | | |
| | | DNase hypersensitive sites | | | 240 | 10 | 19 | | | |
| | | DNase hypersensitive sites, transcription start sites, chromatin states | | | | | | 2 | | |
| | | histone modifications | | | | 6 | | | | |
| | | None | 2 | 8 | 40 | 24 | 23 | 24 | 2 | 8 |
| | | sequence variants | | | | | 48 | | | |
| | | synthetic elements | | | | | 1 | | | |
| | | TF binding sites | | | | 10 | | | | |
| | | transcription start sites | | | | | | 1 | | |
| | in vitro differentiated cells | None | | | 6 | | | 2 | | |
| | organoid | accessible genome regions, histone modi.. | | | | | | 2 | | |
| | | accessible genome regions, histone modifications, DNase hypersensitive sites | | | | | | 2 | | |
| | primary cell | None | | | | | 3 | | | |
| | | sequence variants | | | | | 1 | | | |
| *Mus musculus* | cell line | DNase hypersensitive sites | | | 5 | | | | | |
| | | None | | | 5 | | 6 | | | |

Fig 3: Plot demonstrating the diversity of ENCODE Functional characterization experiments classified by organism, biosample classification, element selection method and various assays.

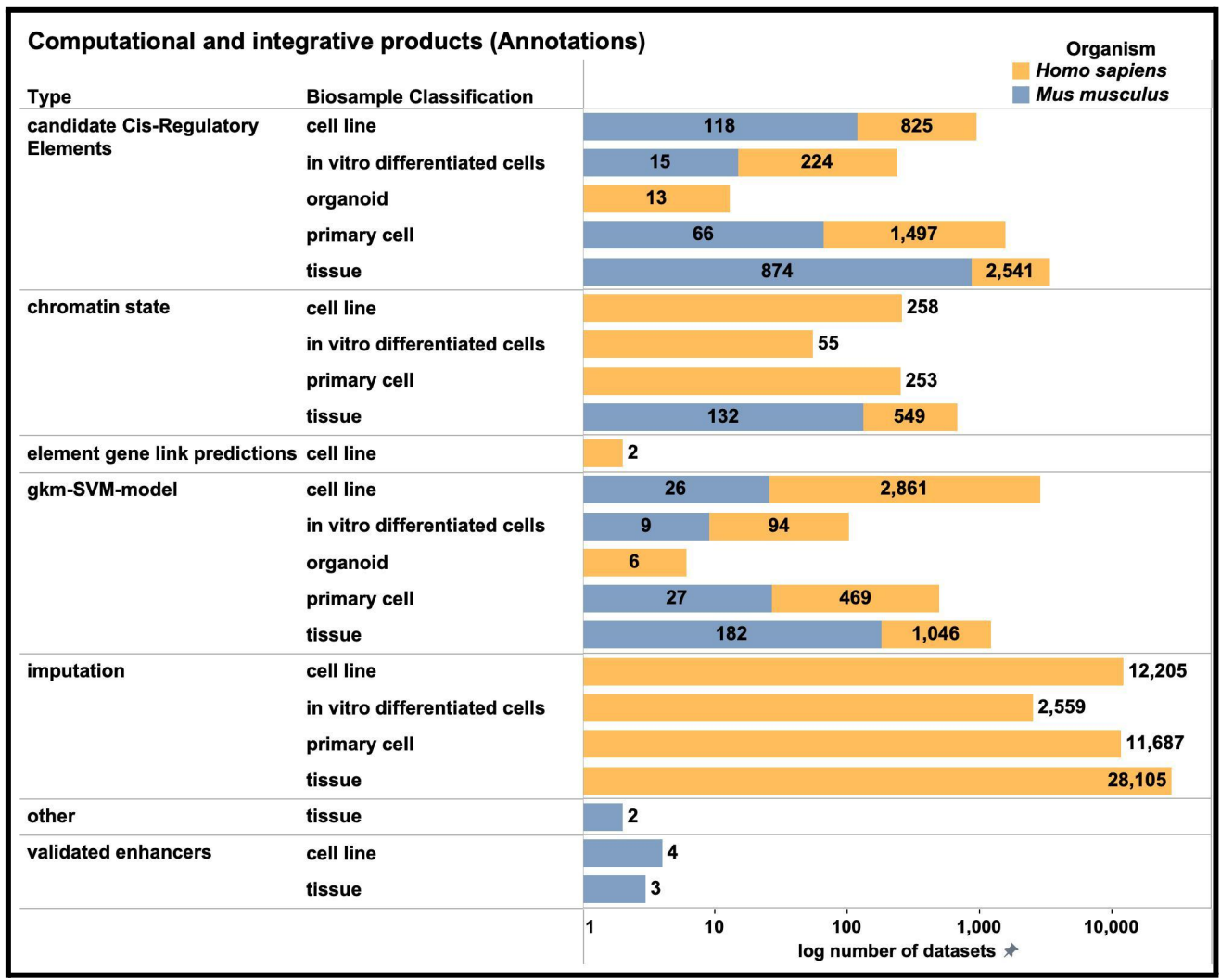

Fig 4: Plot demonstrating the number of computational and integrative products (annotation datasets) classified by organism, biosample classification and annotation type.

| Collection | Organism | 3D chromatin structure | DNA accessibility | DNA binding, Histone | DNA binding, TF | DNA methylation | DNA sequencing | Genotyping | Proteomics | Replication timing | RNA binding | Single cell, DNA accessibility | Single cell, Transcription | Transcription | Total datasets |
|---|---|---|---|---|---|---|---|---|---|---|---|---|---|---|---|
| Deeply profiled cell lines | Homo sapiens | 29 | 13 | | | | | | | | | 15 | | 133 | 190 |
| Degron | Homo sapiens | 45 | 31 | 92 | 16 | | | | | | | 6 | | 62 | 252 |
| ENCORE | Synthetic constructs | | | | | | | | | | 213 | | | | 213 |
| | Homo sapiens | | | | 68 | | | | | | 255 | | | 1,009 | 1,332 |
| ENTEx | Homo sapiens | 8 | 119 | 441 | 242 | 127 | 12 | 4 | | | 2 | 45 | | 276 | 1,276 |
| H1 stem cells | Homo sapiens | 4 | 8 | 136 | 87 | 8 | | 1 | 3 | 1 | 1 | | | 64 | 313 |
| H9 stem cells | Homo sapiens | 3 | 20 | 201 | 23 | 8 | | 1 | | 8 | | 8 | | 41 | 313 |
| Human Donor | Homo sapiens | 163 | 1,444 | 2,746 | 456 | 288 | 12 | 44 | | 17 | 2 | 264 | 129 | 1,379 | 6,944 |
| Human reference epigenomes | Homo sapiens | | 288 | 2,056 | 283 | 91 | | | | | | | | 342 | 3,060 |
| Immune cells | Homo sapiens | 64 | 509 | 948 | 15 | 7 | | | | | 1 | 28 | 1 | 163 | 1,736 |
| Mouse development | Mus musculus | | 305 | 597 | 20 | 84 | | | | | | 48 | 331 | 471 | 1,856 |
| Mouse reference epigenomes | Mus musculus | | 108 | 669 | 62 | 72 | | | | | | | | 175 | 1,086 |
| PGP stem cells | Homo sapiens | 3 | 10 | 73 | 29 | 5 | 2 | 1 | | | | 4 | | 29 | 156 |
| Rush Alzheimers Disease Study | Homo sapiens | 3 | 190 | 176 | 59 | | | | | | | | | 172 | 600 |

**Fig 5:** Number of functional genomics experiments represented within various ENCODE collections classified by assay type and organism. *Note: the numbers reflected above exclude control datasets.*

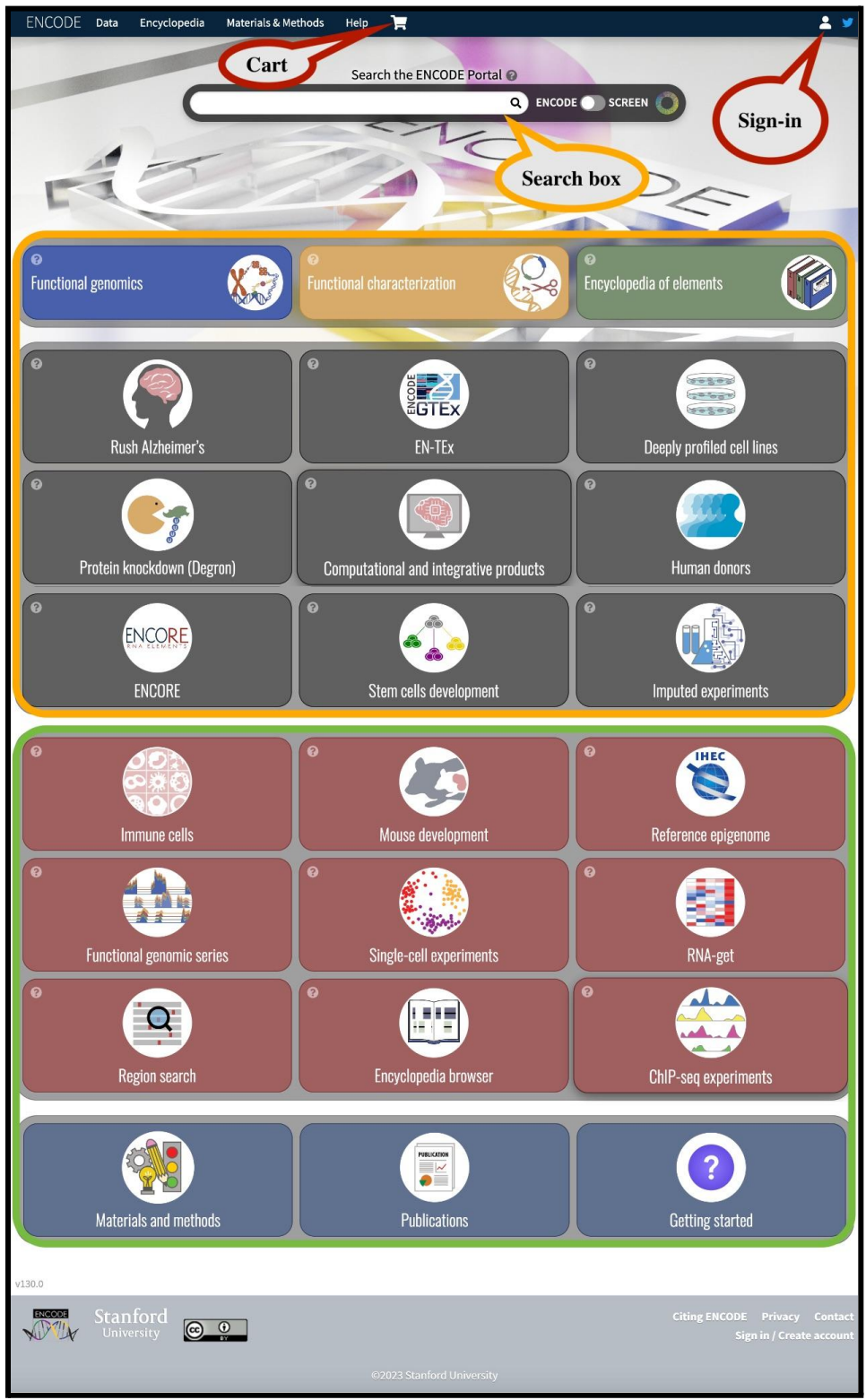

**Fig 6:** ENCODE home page showing different clickable shortcut cards, search box, cart feature and sign-in button.

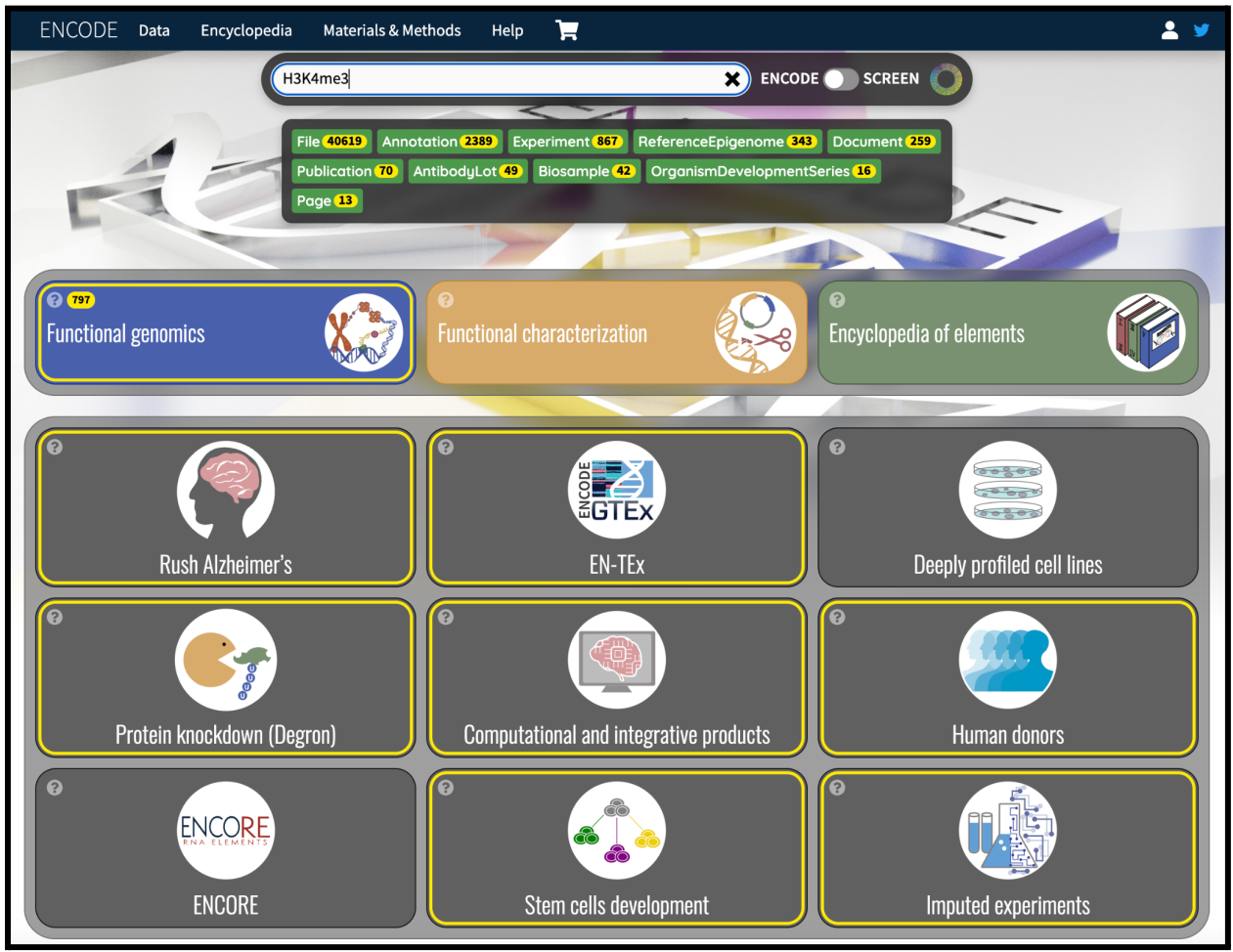

**Fig 7:** Search box usage example. Searching H3K4me3 in the search box highlights relevant homepage cards that include any data related to the searched term. In this case out of the 12 cards, 8 cards are highlighted yellow indicating presence of relevant data within those respective cards.